\documentclass[twoside]{ilcws07}
\usepackage[latin1]{inputenc}
\usepackage[dvips]{graphicx,epsfig,color}
\usepackage{wrapfig,rotating}
\usepackage{amssymb,amsmath,array}

\begin{document}
\title{
Resolution Studies of GEM/Timepix Detector with 5 GeV electrons} 
\author{A. Bamberger$^1$, K. Desch$^2$, U. Renz$^1$, M. Titov$^{1,3}$, N. Vlasov$^1$,
P. Wienemann$^2$, \\
S. Zimmermann$^1$ and A. Zwerger$^1$
\vspace{.3cm}\\
1- Albert-Ludwigs University of Freiburg, Physics Institute, Freiburg, Germany
\vspace{.1cm}\\
2- Rheinische Friedrichs-Wilhelms-Universit\"at, Physics Institute, Bonn, Germany
\vspace{.1cm}\\
3- CEA Saclay, DAPNIA/SPP, Gif sur Yvette, France
}

\maketitle

\begin{abstract}
This contribution investigates a prototype of a TPC readout with a highly pixelated CMOS ASIC,
which is an option for charged particles tracking of the ILC.
A triple GEM stack was joined with a TimePix and MediPix2 chip (pixel size of 55$\times$55\,$\mu m^2$)
and its readout properties were investigated with 5 GeV electrons.
The spatial resolution of the cluster center reconstruction was
determined as a function of drift distance using different cluster alhoritms 
and compared with Monte Carlo predictions.
\end{abstract}

\section{General consideration}

The recent development of Micro-Pattern-Gas-Detectors (MPGD) allows an 
extended field of application for detectors with gas multiplication. 
For Time-Projection-Chambers (TPC) the readout with Gas-Electron-Multipliers (GEM) 
\cite{sauli} has been demonstrated in conjunction with an unconventional
readout option with high pixelation using the MediPix2 chip \cite{pub}. 
The TimePix~\cite{Xavier} and Medi\-Pix2 is investigated the first time with 
5 GeV electrons from DESY II. They have the virtue of negligible multiple 
scattering. An external tracking with Si-telescope provides information on the track
position which is used for resolution determination and for the drift
velocity measurements using a TimePix chip.

\section{Basic setup for the GEMs and the MediPix2/TimePix chip}
 \begin{figure}[h]
  \psfig{figure=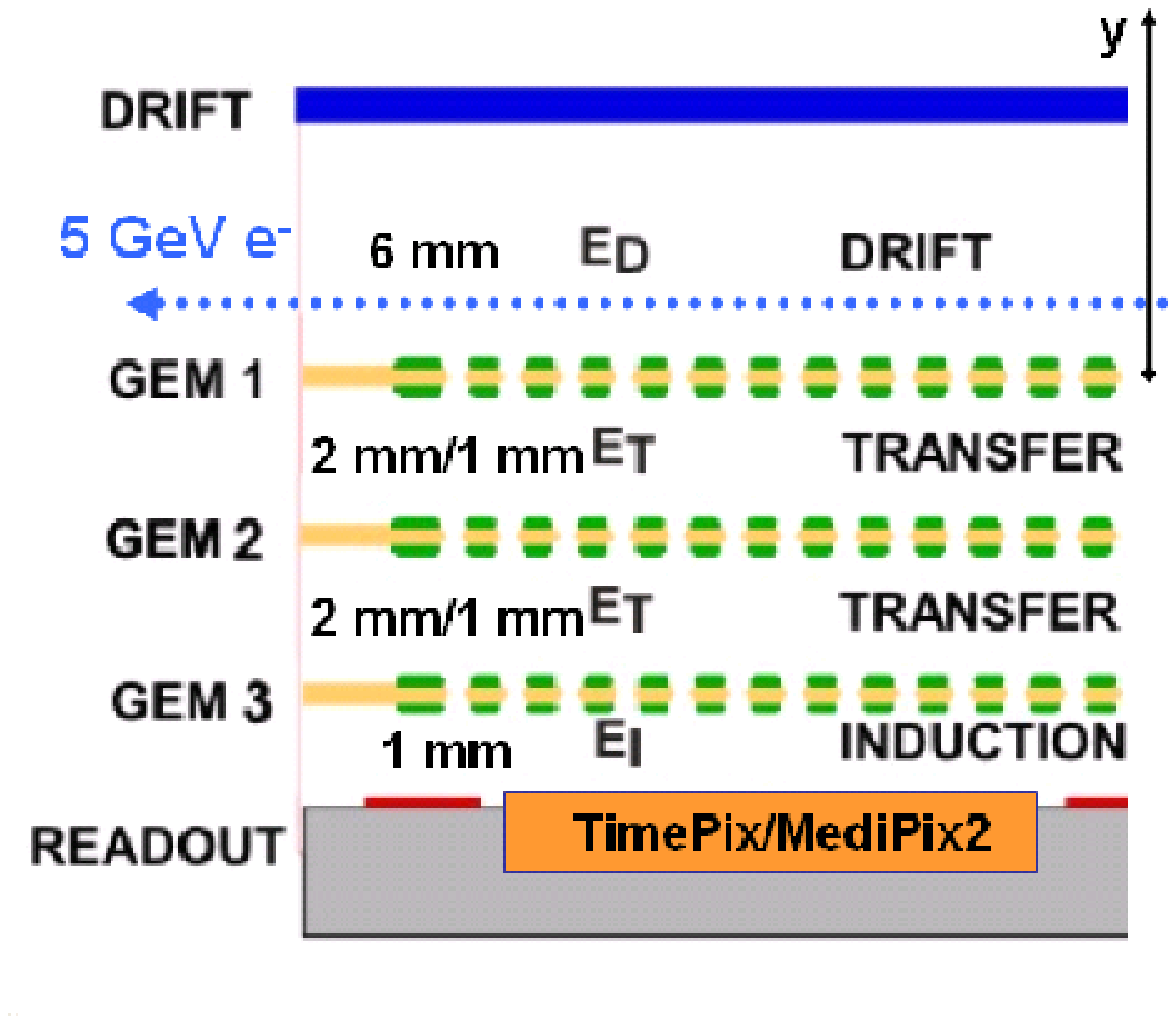,width=0.5\textwidth} \hskip -0.5cm
  \psfig{figure=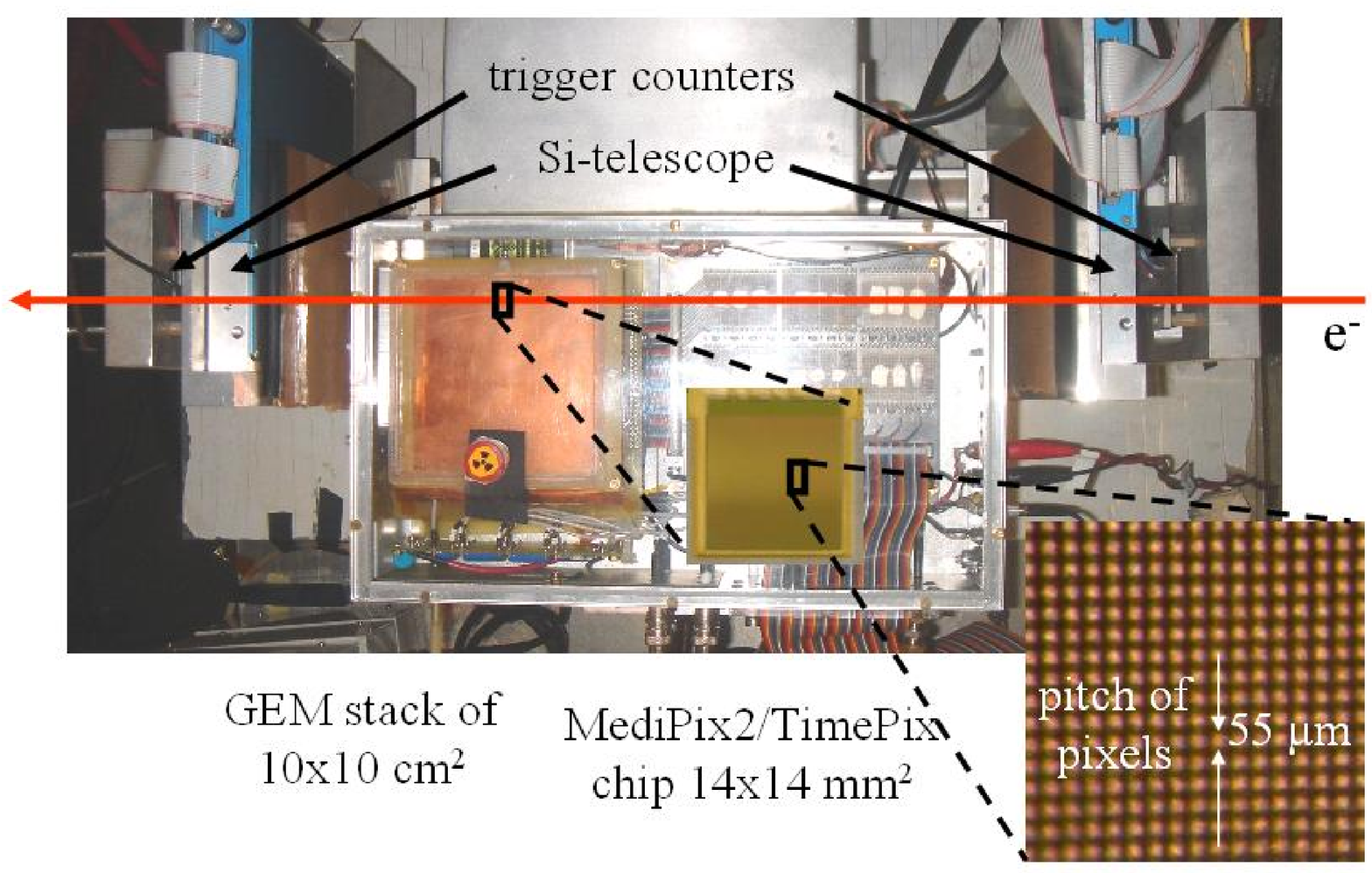,width=0.55\textwidth}
\label{stack}
\caption{Setup of drift volume: stack and MediPix/TimePix (left) and
 test beam setup where inserts show size of MediPix2/TimePix and the size of a pixel (right)}
  \end{figure}

A cross sectional view of the triple GEM plus Medipix2 and Timepix detector
with an electron beam crossing the drift region is shown in Fig.1 (left part).
The drift volume of 6 mm thickness and 10$\times$10 cm$^2$ in size serves for charged track detection. 
The drift field is about 1.1 kV/cm. 
Three CERN-produced GEMs of the same area as the drift volume with 70\,$\mu$m holes
of 140\,$\mu$m pitch are arranged in a stack above the readout plane with 2\,mm (2-2-1)
or 1\,mm distances (1-1-1) for the transfer gaps, the latter provides a collapsed setup
in order to study possible effects on the spatial resolution.  
With the triple GEM stack gas amplification of $\approx$10$^5$ can be achieved with
Ar/CO$_2$ and He/CO$_2$\,-\,mixtures for the high resolution detection of minimum ionizing particles. 
The MediPix2 and TimePix chip is positioned at a distance 
of 1\,mm from the last GEM exposed to a field of E$_I$ = 4.0 kV/cm. 
The advantage of a GEM setup is the robust operation, since the fields of the gas
multiplication region ($\approx$ 70 kV/cm) are well shielded insige GEM holes, see \cite{pub}.

\section{The test beam setup}

The MediPix2 and Timepix chip has a surface of 14$\times$14 mm$^2$ and a pixel size of 55$\times$55\,$\mu m^2$.
It is positioned close to the border of the GEM stack, see Fig. 1 (right part). 
The remaining surface of the GEMs is covered with 24 anode pads of 
2$\times 2$\,cm$^2$ size for monitoring purposes.
The gas tight box contains the GEMs, the resistor chain, the TimePix and MediPix2 chip
and the readout electronics of the pads.\\
The readout of the MediPix2 and TimePix is done with MUROS2 using the fast shutter option.
The thresholds for the pixels used were 990\,e$^-$ and 830\,e$^-$ for the
MediPix2 and the TimePix, respectively. 
The electron beam is defined by trigger scintillating counters of 1$\times$1.5 cm$^2$ in size
and a Si-telescope with 3 planes of strips (two planes allow measurement of 
the x-coordinate in front and behind the GEM plus Medipix2 and Timepix, and one plane is used for the
y-coordinate measurement in front of the detector).
The effective readout pitch of Si-telescope is 50\,$\mu$m.   

\section{TimePix: The TIME, TOT and MIXED modes}

The TimePix has a clock, which is distributed throughout the entire chip.
A register on each pixel counts the number of
clock cycles in a way depending on the chosen mode of operation. For each pixel this mode
can be set individually. In TIME-mode the cycles are counted from the point when the signal
crosses the threshold till a common stop by the gate signal
(=\,"Fast shutter"). The other mode "Time-Over-Threshold" records the clock cycles as
long as the pulse is above the threshold. The maximum number of counts in this measurement
is limited by the chosen gate width of $12.6\,\mathrm{\mu m}$,
which is about 600 counts at a given clock frequency of 48\,MHz.

In a special configuration ``Mixed Mode'' every other pixel switches  
the TIME and TOT mode in checker board fashion. This results in 1/2 of all pixels
are of TIME-type and another 1/2 of the TOT-type.
This allows a proximity information of both TIME and TOT. 

\section{Cluster reconstruction and point resolution}

Typically, 8 - 9 non-separated clusters per track were observed, see
Fig. 2\, (left part). Images with double tracks were not
considered in the current analysis. The following three clustering methods
were used, where last two can separate the overlapping clusters using the charge
deposition information (TOT or MIXED modes) resulting in everage 10 - 11 clusters
per event:
\begin{itemize}
\item ``Contiguous areas'' method leaves overlapping clusters not separated. That
 method is applicable for any TimePix mode. The result is a lower number of clusters
 depending on the effective threshold settings and therefore not covered here
\item ``Saddle Point'' method separates the contiguous clusters making a line
for secondary maximum in a projection transverse to track. The line divides
the merged clusters at the saddle point
\item ``Island'' method joins the adjacent pixels with a nonzero TOT values into
clusters in the way that a pixel is joined into the cluster of its neighbor with
a highest TOT value. The procedure is repeated for each pixel to produce a unique
assignment of pixels to clusters. It results in a separation in two spacial coordinates
available compared with the above case
\end{itemize}
 \begin{figure}[h]
  \psfig{figure=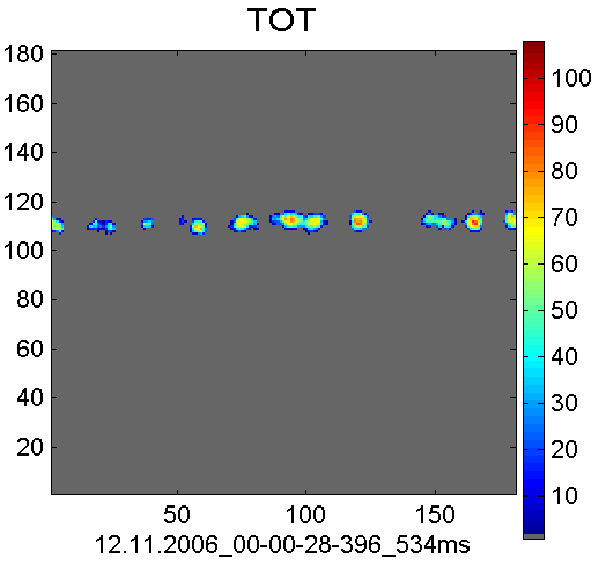,width=0.45\textwidth} \hskip -0.1cm
  \psfig{figure=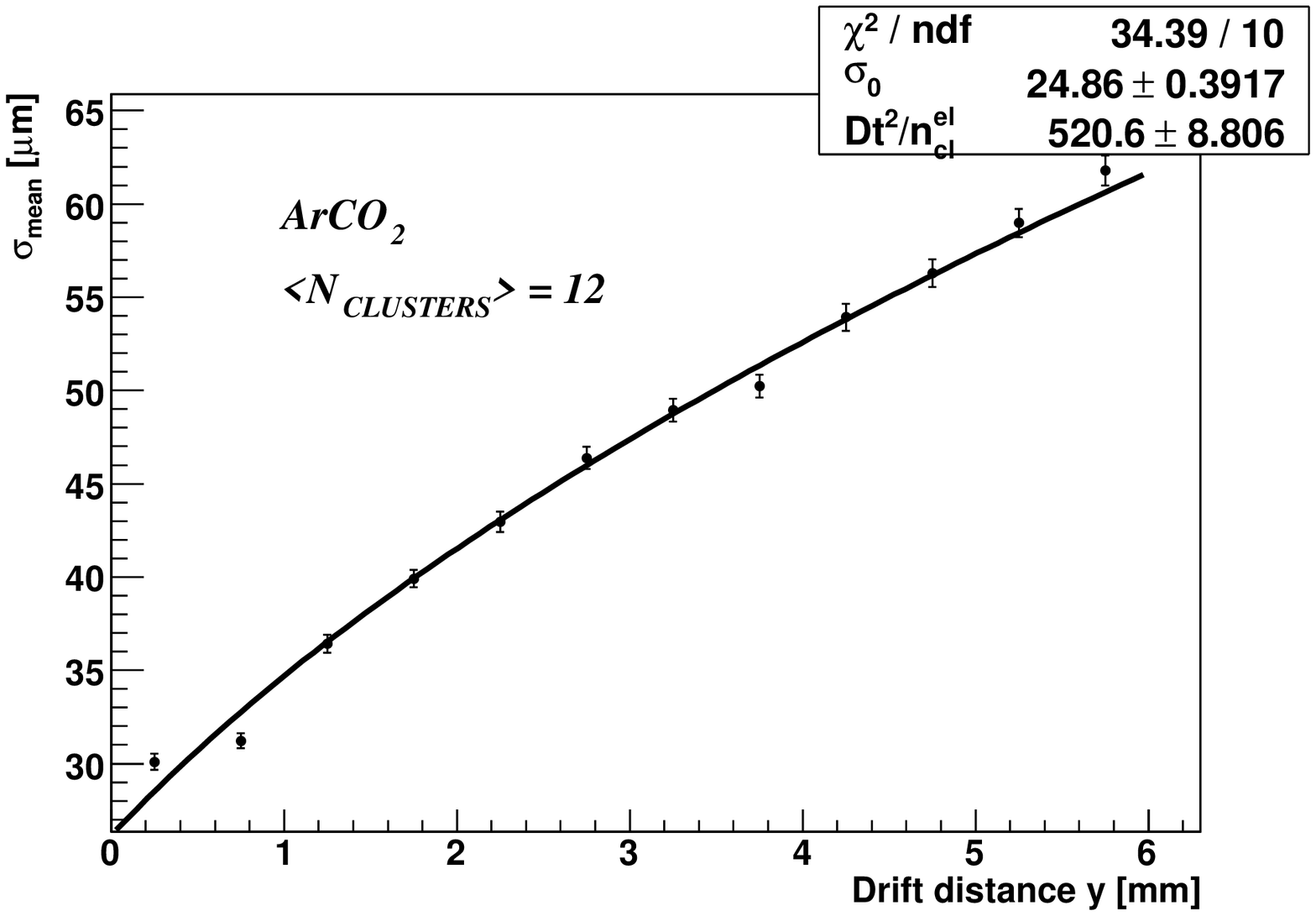,width=0.55\textwidth}
\label{sigma0}
\caption{Image of 5\,GeV electron track (left part) and $\sigma_0$ and $D_t^2/n_{cl}^{el}$ for Ar/CO$_2$ 
(right part)}
  \end{figure}

The evaluation follows basically \cite{pub}: A cluster is rejected if it
contains less than 9 hits. The noise is removed beyond the cluster region.   
Two different methods of the straight line fits to the centers of
the clusters result in an inbiased estimation of the standard deviation.
Residuals of the fit to all cluster centroids (N) and a fit to (N-1) 
centroids resulting in a residual of the exempt cluster which is
permutated over all clusters. These residuals enter into two Gaussian fits which
produce $\sigma_{unbiased}=\sqrt{\sigma_{N}\times\sigma_{N-1}}$.
 
The left part of Fig.~2 shows a measured dependance of the resolutions versus the drift space for Ar/CO$_2$.
In the drift region to separate the lateral diffusion within the drift space from the intrinsic
GEM plus TimePix and MediPix2 resolution the y-coordinate information of the external telescope is used.
The following parametrisation is applied for the fit: $\sigma^2=\sigma_0^2+\frac{D_t^2\times y}{n_{cl}^{el}}$.
The $\sigma_0$ equal to 24.9 $\pm$ 0.4 results from the fit for Ar/CO$_2$.

\section{Comparison with Monte Carlo simulations}

The results of spatial resolution $\sigma_0$ and $D_t^2/n_{cl}^{el}$ corresponding to different
clustering methods for Ar/CO$_2$ and He/CO$_2$ are summarised in Tab.~2.
The comparison with HEED simulations~\cite{hausch} is also given in the table.

\begin{table}[h]
\begin{tabular}{|c|c|c|c|c|c|}
\hline 
\multicolumn{6}{|c|} {~~~~~~~~~~~~~~~~~~~~~~~~~~~~~~DATA~~~~~~~~~~~~~~~~~~~~~~Simulations}\\ \hline
    & Gas &  $\sigma_0$ & $D_t^2/n_{cl}^{el}$ & $\sigma_0$ & $D_t^2/n_{cl}^{el}$ \\ \hline  \hline
 ``Island''   & Ar/CO$_2$ & 24.9 $\pm$ 0.4 & 521 $\pm$ 9 & -- & -- \\
              & He/CO$_2$ & 29.4 $\pm$ 0.5 & 660 $\pm$ 14 & -- & -- \\\hline
  ``Saddle    & Ar/CO$_2$ & 18.4 $\pm$ 2.7  & 467 $\pm$ 36 & 15.2 $\pm$ 3.8 & 726 $\pm$ 41 \\
    Point''   & He/CO$_2$ & 27.1 $\pm$ 4.9  & 547 $\pm$ 78 & 19.4 $\pm$ 4.0 & 989 $\pm$ 54 \\
\hline 
\end{tabular}
\label{table}
\caption{The values of $\sigma_0$ and $D_t^2/n_{cl}^{el}$ for the data and Monte Carlo simulations.
The comparison with Monte Carlo simulation
is present at the moment only for the ``Saddle Point'' clustering method.}
\end{table}

The agreement between two clustering methods is found to be good for both
$\sigma_0$ and $D_t^2/n_{cl}^{el}$ values.
It is found that simulations considerably overestimate the slope $D_t^2/n_{cl}^{el}$ comparing
to the measurements for both gases.

The slope depends on the size of merged clusters because a higher number of electrons per
detected cluster than for an ideally resolved cluster is present. This leads to a reduction 
of the effective transverse diffusion through the active volume and reduces the slope parameter.
Based on this observation it is possible that the discrepancy of the Monte Carlo with
respect to the slope is present, since the average number of detected clusters lower 
by a factor of 2 than expected from simulations~\cite{hausch}.

\section{Acknowledgments}
We would like to thank the EUDET project for its financial support and the MediPix collaboration
for supporting us work with readout software and hardware, especially with respect to the
TimePix chip used for the very first time in an experiment. We
would also like to thank Michael Campbell, Erik Heijne, Xavier Llopart and
Fabio Sauli for stimulating discussions and a lot of valuable advices.

\begin{footnotesize}

\end{footnotesize}


\end{document}